\documentclass[twocolumn,showpacs,preprintnumbers,amsmath,amssymb,floatfix,pra,aps,10pt,letterpaper]{revtex4-1}
\usepackage{graphicx}
\usepackage{dcolumn}
\usepackage{bm}
\usepackage{amsmath}
\usepackage{amsfonts}
\usepackage{mathrsfs}

\begin{document}

\title{Formation of H$_3^-$ by radiative association of H$_2$ and H$^-$ in the interstellar medium}
\author{ M. Ayouz$^{1}$,   R. Lopes$^{1}$, M. Raoult$^{1}$, O. Dulieu$^{1}$, and V. Kokoouline$^{1,2}$}
\affiliation{$^{1}$Laboratoire Aim\'e Cotton, CNRS, B\^at 505, Universit\'e Paris 11, 91405 Orsay Cedex, France\\$^{2}$Department of Physics, University of Central Florida, Orlando, Florida 32816, USA }


\begin{abstract}
We develop the theory of radiative association of an atom and a diatomic molecule within a close-coupling framework. We apply it to the formation of H$_3^-$ after the low energy collision (below 0.5~eV) of H$_2$ with H$^-$.  Using recently obtained potential energy and permanent dipole moment surfaces of H$_3^-$, we calculate the lowest rovibrational levels of the H$_3^-$ electronic ground state, and the cross section for the formation of H$_3^-$ by radiative association between H$^-$ and ortho- and para-H$_2$. We discuss the possibility for the H$_3^-$ ion to be formed and observed in the cold and dense interstellar medium in an environment with a high ionization rate. Such an observation would be a probe for the presence of H$^-$ in the interstellar medium.
\end{abstract}


\maketitle

\section{Introduction}
\label{sec:intro}

Many chemical reactions in the interstellar medium (ISM) are powered by cosmic rays: atoms and molecules (mainly molecular hydrogen) are ionized by radiation that provides sufficient energy to initiate a chain of chemical reactions in interstellar clouds leading to the synthesis of polyatomic molecules. About fourteen positive ions have been observed and identified in the ISM, including the simplest triatomic one, H$_3^+$. This ion plays an important role in chemistry and evolution of interstellar clouds \cite{geballe96,oka06b}, as its abundance is strongly related to the production of H$_2^+$ in the ISM. In contrast, the existence of stable negative ions has long been thought impossible due to the presence of UV radiation in the ISM. In particular, the H$^-$ ion, which has a single bound state, has not been directly detected yet so that its presence, though probable, is still controversial \cite{ross08}. Even if it has been proposed some time ago \cite{herbst81}, the surprise came with the detection of negative molecular ions in dense molecular clouds or carbon star envelopes. There are several carbon chain ions that have been detected so far: C$_6$H$^-$ \cite{mccarthy06,cernicharo07,cernicharo08}, C$_4$H$^-$ \cite{gupta07}, C$_8$H$^-$ \cite{kawaguchi07}, C$_3$N$^-$ \cite{thaddeus08}, C$_5$N$^-$ \cite{cernicharo07,cernicharo08}, and CN$^-$ \cite{agundez10}. All these species are closed-shell systems with a quite large electron binding energy (3.8~eV for C$_6$H$^-$), and a large permanent dipole moment of several debye. Although it is generally accepted that in the ISM these ions are formed after radiative capture of an electron by the parent neutral molecules \cite{herbst08}, no detailed quantum mechanical treatment has been developed yet for modeling the process.

As the negatively charged counterpart of H$_3^+$, the  H$_3^-$ ion is predicted to be stable by about 0.013~eV \cite{starck93,ayouz10}. It has never been detected so far in the cold regions of the ISM, while Wang \textit{et al.} \cite{wang03} observed it by mass spectrometry of laboratory plasmas. The H$_3^-$ ion cannot be formed by radiative attachment from the unstable H$_3$ molecule, so that one has to consider alternative paths: the three-body recombination (TBR) $\mathrm{H}_2+\mathrm{H}^-+\mathrm{X} \to  \mathrm{H}_3^- +\mathrm{X}\,$, or the radiative association (RA) $\mathrm{H}_2+\mathrm{H}^- \to  \mathrm{H}_3^- +\hbar\omega\,$.
The TBR mechanism must be the dominant process of H$_3^-$ formation in the laboratory plasma but it is most probably inefficient in the ISM because of low H$_2$ densities. The goal of this study is to investigate the possibility to form H$_3^-$ by the radiative association of H$_2$ and H$^-$ in low temperature ($<$150~K) environments.

The collisions between H$^-$ and H$_2$ have been studied both theoretically \cite{starck93,panda04} and experimentally \cite{muller96,wester09}, only for collision energy larger than 0.5~eV. In contrast, the structure of the H$_3^-$ ion has been rarely explored in the past \cite{starck93}. Recently, we have calculated  (see Ref. \cite{ayouz10}, hereafter referred to as paper I) a new accurate potential energy surface (PES) for the H$_3^-$ electronic ground state, whose accuracy was improved compared to the previous {\it ab initio} calculations \cite{starck93,panda04}. In addition, we have determined for the first time the H$_3^-$ permanent dipole moment surface (PDMS), which is needed for the RA calculations. The H$_3^-$ ion is well represented as a loosely-bound H$_2 \cdots $H$^-$ complex with several rovibrational states, bound at most by about 70~cm$^{-1}$. We have also found that there are a number of predissociation resonances, which can be described as excited rovibrational states $(j,v)$ of H$_2$ perturbed by H$^-$, coupled to the dissociation continuum H$_2(j',v')$+H$^-$ with energy of the dimer state $E(j',v')$ lower than $E(j,v)$.

The paper is organized as follows. In the next section we discuss the geometry of the H$_3^-$ molecule.  In section \ref{sec:wf} we introduce the rovibrational wave functions of the molecule for bound and continuum states. The theory of radiative association of a dimer and an atom is developed in section \ref{sec:RA-theory}. Finally, in section \ref{sec:ism}  we present results of numerical calculations of the RA cross-section and of the rate coefficient for H$_3^-$ formation, and we discuss the possibility to observe the H$_3^-$ ion in the ISM.

\section{Representation of the H$_2 \cdots $H$^-$ complex at low energies}
\label{sec:sym}

The H$_3^-$  molecule is composed of three identical nuclei, described in principle within the CNPI (complete nuclear permutation inversion) group $D_{3h}$ \cite{bunker98}. However, due to the loosely-bound nature of the H$_3^-$ ion in its electronic ground state \cite{starck93,ayouz10}, the exchange probability of the H$^-$ proton with the dimer protons is negligible. Therefore, the low-energy collision between H$_2$ and H$^-$ can be studied as an inelastic collision of two structured particles. We first define two coordinate systems: The space-fixed (SF) frame with axes $(x,y,z)$, and the body-fixed (BF) frame with axes $(X,Y,Z)$ whose orientation in the SF frame is given by the Euler angles $\alpha_e$, $\beta_e$ and $\gamma_e$. The $Z$-axis connects the center of H$_2$ molecular axis with the nucleus of H$^-$, and its $X$-axis is in the plane of the three nuclei. The orientation of the dimer with respect to $Z$ is given by the azimuthal angle $\theta$. The natural coordinate system (CS) associated to the BF frame is the Jacobi CS $(R,r,\theta)$ where $R$ is the distance between H$_2$ and H$^-$ along the $Z$ axis, and $r$ the internuclear distance of H$_2$.

\begin{figure}
\includegraphics[width=7.cm]{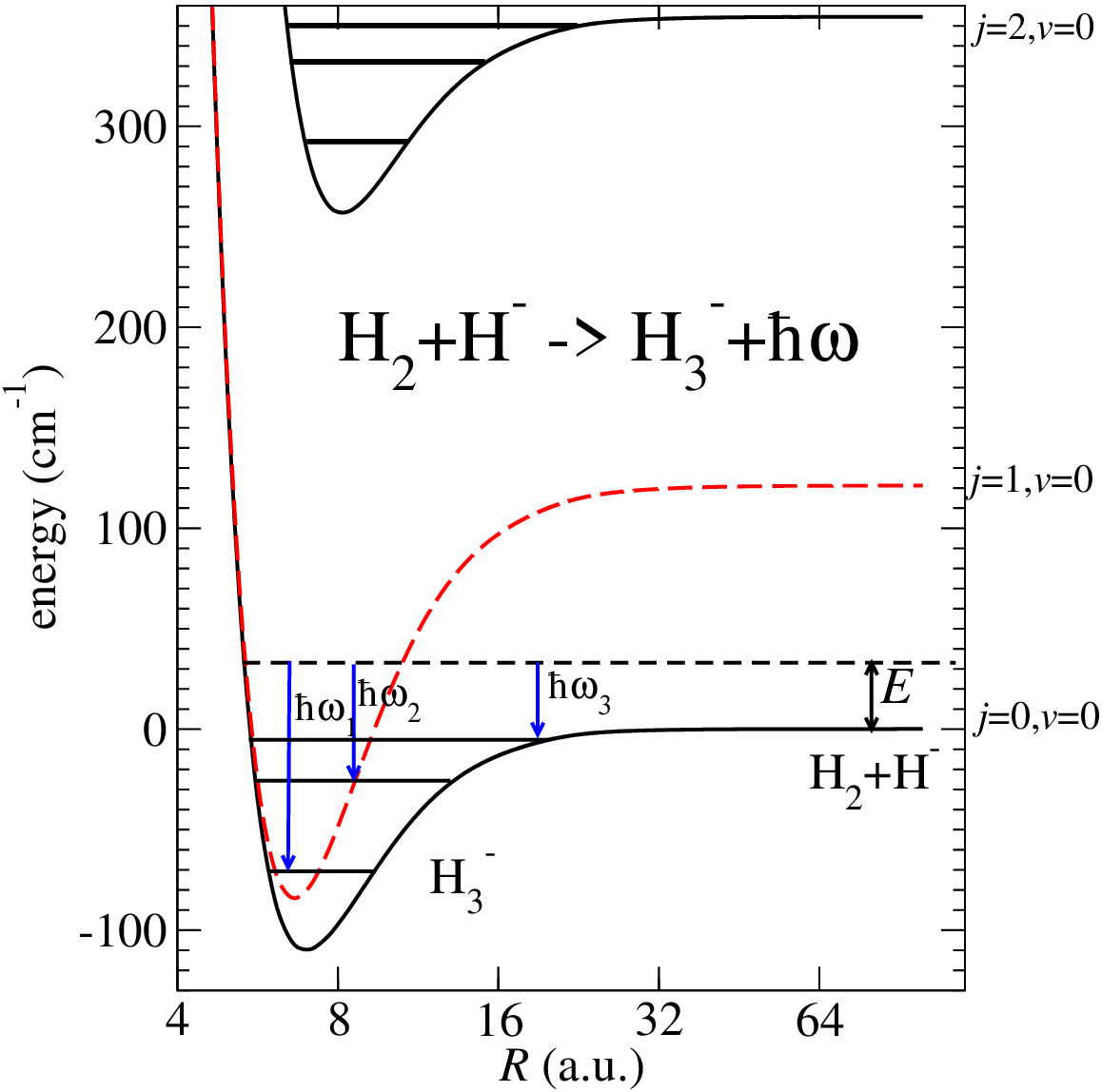}
\caption{(Color online)  The adiabatic curves of H$_3^-$ as functions of the Jacobi coordinate $R$, and the scheme for radiative association at a low collisional energy $E$.  Horizontal lines indicate positions of several bound states (reported in Table \ref{tab:para_bound_states}) and resonances.}
\label{fig:ad_curves}
\end{figure}

In the framework of this "super-dimer" approximation, the quantum numbers of the H$_2$ dimer are treated as good quantum numbers. The quantum states of the H$_2 \cdots $H$^-$ complex at low energies are characterized by the quantum numbers $(J,M,j,v,\Omega^\pm)$: $J$ and $j$, associated to the total angular momentum $\widehat{\textbf{J}}$ of the complex and to the angular momentum $\widehat{\textbf{j}}$ of the dimer, the vibrational state $v$ of H$_2$, the absolute values $M$ and $\Omega$ of the projection of $\widehat{\textbf{J}}$ on the SF $z$ axis, and of $\widehat{\textbf{j}}$ on the BF $Z$ axis, and the intrinsic parity $\pm$ of the wave function with respect to the reflection $\sigma_v$ through the plane containing the  $Z$-axis. For $\Omega\ne 0$ the $\Omega^+$ and $\Omega^-$ states are degenerate in this approximation. If the H$_2 \cdots $H$^-$ complex is bound, an additional quantum number $v_t$ characterizes the vibration along the $Z$ axis. One defines the basis  set $|J,v,j,\Omega \rangle$
\begin{eqnarray}
\label{eq:bfbasis}
|J,v,j,\Omega \rangle = \\
\sqrt{\frac{2J+1}{4\pi}}\left[ D_{M 0}^{J}(\alpha_e,\beta_e,\gamma_e)\right]^*Y_{j\Omega}(\theta,\gamma) \chi_{vj}(r) \nonumber\,,
\end{eqnarray}
or, in the equivalent form
\begin{eqnarray}
\label{eq:bfbasis2}
|J,v,j,\Omega \rangle = \\
\sqrt{\frac{2J+1}{8\pi^2}}\left[ D_{M\Omega}^{J}(\alpha_e,\beta_e,\gamma_e)\right]^*\Theta_{j}^{\Omega}(\cos\theta) \chi_{vj}(r) \nonumber\,,
\end{eqnarray}
where the normalized function $\Theta_{j}^{\Omega}(\cos\theta)=P_{j}^{\Omega}(\cos\theta)/\sqrt{4\pi}$ \cite{landau3} is proportional to the associated Legendre polynomial and describes the rotational state of the dimer, and $\chi_{vj}(r)$ is the vibrational wavefunction of H$_2$. The Wigner function $D_{M\Omega}^{J}(\alpha_e,\beta_e,\gamma_e)$ is associated to the rotation of the complex in the SF frame. The function above is not yet symmetrized with respect to $\sigma_v$ and, correspondingly, $\Omega$ can be positive or negative. Assuming that the PES $V(R, r, \theta)$ is known (see paper I), one derives the interaction matrix \textbf{V} (diagonal with respect to $\Omega$) written in the above basis (after integration over $r$ and $\theta$, denoted by the subscripts of the angled brackets)
\begin{equation}
\label{eq:bfpot}
 V_{vj\Omega,v'j'\Omega'}^{J}(R)=\langle Jvj\Omega |V(R,r,\theta)|J v'j'\Omega\rangle_{r,\theta} \delta_{\Omega \Omega'}\,.
\end{equation}
As $R \rightarrow \infty$, the diagonal (in $v$ and $j$) elements of this matrix become equal to the energies $\epsilon_{vj}$ of the rovibrational quantum states $|vj\rangle$ of H$_{2}$. The diagonalization of the \textbf{V} matrix at every $R$ value yields a set of adiabatic potential energy curves, correlated at large $R$ to the sum of $\epsilon_{vj}$ and of the H$^-$ ground state (Fig. \ref{fig:ad_curves}). These curves can accommodate bound levels corresponding to the quantization of the vibrational motion along the $R$ coordinate. The H$_2$ nuclear spin  $i_d=0$ (para-H$_2$) or $i_d=1$ (ortho-H$_2$) determines the possible values of $j$, namely even in the former case, and odd in the latter case. The energy of the state does not depend on the quantum number $M$ which will be omitted in the following.

The wave functions of Eq. (\ref{eq:bfbasis2}) must be properly symmetrized with respect to $\sigma_v$ and total inversion $E^*$, corresponding to the symmetry operators of the $C_{\infty v}$ group relevant to the present approximation. The $|J,v,j,\Omega\rangle$ functions obey the transformations
\begin{eqnarray}
\label{eq:sigmav}
\sigma_v|J,v,j,\Omega\rangle=|J,v,j,-\Omega\rangle\nonumber\,,\\
E^*|J,v,j,\Omega\rangle=(-1)^J|J,v,j,-\Omega\rangle\,,
\end{eqnarray}
so that we define the appropriate symmetrized combinations
\begin{eqnarray}
\label{eq:sigmav_sym}
|J,v,j,\Omega^\pm\rangle=\left(|J,v,j,\Omega,\rangle\pm|J,v,j,-\Omega\rangle\right)/\sqrt{2}\,.
\end{eqnarray}
From these equations, the total parity of the $+$ and $-$ states is given by the value of $(-1)^{J}$ and $(-1)^{J+1}$, respectively (also referred to as the \textit{e} and \textit{f} parity for linear molecules \cite{brown1975}). The permutation $(12)$ of the the two nuclei of H$_2$ requires another symmetrization of the dimer wave functions
\begin{equation}
\label{eq:(12)}
(12)|J,v,j,\Omega\rangle=(-1)^j|J,v,j,\Omega\rangle \,,
\end{equation}
so that we can write the following four wave functions labeled by the irreducible representations (irreps) of the $C_{2v}$ group $\Gamma=A_1,A_2,B_1,B_2$ (see Eq. (94.18) of \cite{landau3}) (with $\Omega \ge 0$):
\begin{eqnarray}
\label{eq:C2v_sym}
|A_1,J,v,j,\Omega\rangle={\cal N}(|J,j,\Omega\rangle+(-1)^j|J,v,j,\Omega\rangle \nonumber\\
+(-1)^{J}|J,j,-\Omega\rangle+(-1)^{J+j}|J,v,j,-\Omega\rangle)\\
\label{eq:C2v_sym2}
|A_2,J,v,j,\Omega\rangle={\cal N}(|J,j,\Omega\rangle+(-1)^j|J,v,j,\Omega\rangle \nonumber\\
-(-1)^{J}|J,j,-\Omega\rangle-(-1)^{J+j}|J,v,j,-\Omega\rangle)\\
\label{eq:C2v_sym3}
|B_1,J,v,j,\Omega\rangle={\cal N}(|J,j,\Omega\rangle-(-1)^j|J,v,j,\Omega\rangle \nonumber\\
-(-1)^{J}|J,j,-\Omega\rangle+(-1)^{J+j}|J,v,j,-\Omega\rangle)\\
\label{eq:C2v_sym4}
|B_2,J,v,j,\Omega\rangle={\cal N}(|J,j,\Omega\rangle-(-1)^j|J,v,j,\Omega\rangle \nonumber\\
+(-1)^{J}|J,j,-\Omega\rangle-(-1)^{J+j}|J,v,j,-\Omega\rangle)\,.
\end{eqnarray}
The normalization factor $\cal N$ above is 1/4 or $1/(2\sqrt{2})$ for $\Omega=0$ and  $\Omega > 0$ respectively. Some functions above are identically equal to 0 for certain combinations of $J,j$, and $\Omega$, which means that the corresponding irreps are not allowed to these combinations.

\section{Bound and continuum wave functions of the H$_2 \cdots $H$^-$ complex at low energies}
\label{sec:wf}

The super-dimer approximation above is useful for defining basis functions with the appropriate symmetry. However, the quantum numbers of the H$_2$ internal state can be used at large separation, but not when H$_2$ approaches H$^-$. An accurate description of the wave functions $\vert \Gamma ,J,E\rangle$ of the H$_2 \cdots $H$^-$ complex with energy $E$ ($E$ being either the collision energy, or the bound state energy, of the complex) is obtained within the close-coupling framework for which we recall below the main steps. The close-coupled expansion on $N_{\textrm{tot}}$ symmetrized channel functions $\vert\Gamma,J,a \rangle$ is written as
\begin{equation}
\label{eq:wftot}
\vert \Gamma ,J; E\rangle=\sum_{a=1}^{N_{\textrm{tot}}} C_{a}(E)\vert\Gamma,J,a;E\rangle\,,
\end{equation}
where $a$ is the channel index. The channel function $\vert\Gamma,J,a;E\rangle$ are the independent solutions of the Schr\"odinger equation with appropriate boundary conditions. They are written as a linear combination of $N_{\textrm{tot}}$ basis functions $|\Gamma,J,v,j,\Omega^{\pm} \rangle$
\begin{equation}
\label{eq:bfwf}
\vert \Gamma,J,a;E \rangle=\sum_{k=1}^{N_{\textrm{tot}}} F_{k,a}(R;E)\vert \Gamma,J,v_k,j_k,\Omega_k \rangle \,.
\end{equation}
In these expressions, the index $a$ can be assigned to the asymptotic limit of the channel wave function in the BF frame at large distances, \textit{i.e.} $(a \equiv v_{a},j_{a},\Omega_{a})$. The number of channels $N_{\textrm{tot}}$ is determined by convergence check on the calculated physical property, namely the binding energies or the RA cross sections.

The $F_{k,a}$ expansion coefficients  are solution of the set of coupled differential equations written (in a matrix form) in the BF frame as
\begin{eqnarray}
\label{eq:bfccset}
\left[-\dfrac{\hbar^2}{2m}\dfrac{\partial^{2}}{\partial R^{2}}\textbf{I}+(\boldsymbol{\epsilon}-E \textbf{I}) +\textbf{W}(R)\right] \textbf{F}(R)=0\,,
\end{eqnarray}
where $m$ is the reduced mass of the complex, and $\boldsymbol{\epsilon}$ the diagonal matrix of the H$_2$ energies $\epsilon_{vj}$. The interaction matrix \textbf{W} is the sum of the interaction potential matrix \textbf{V} (Eq. (\ref{eq:bfpot}) and of the matrix \textbf{R} (diagonal in $v$ and $j$) describing the atom-diatom relative rotation with the orbital angular momentum operator $\widehat{\boldsymbol{ l}} = \widehat{\textbf{J}}-\widehat{\textbf{j}}$,
\begin{eqnarray}
\label{eq:bfl}
{R}_{vj\Omega,v'j'\Omega'}^{\Gamma J}&=&\langle \Gamma,J,v,j,\Omega |\widehat{\boldsymbol{ l}}^2|\Gamma,J,v',j',\Omega'\rangle~\dfrac{\delta_{vj,v'j'}}{2m R^2}\,.
\end{eqnarray}
A detailed expression of these matrix elements in the BF frame can be found in Ref.\cite{launay1976}. The operator  $\boldsymbol{\hat  l}^2$ couples channels with different values of $\Omega$, through long-range terms which dominate the potential energy terms due to their $R^{-2}$ character. Therefore in order to treat conveniently a scattering process, it is usual to choose the representation in the SF frame where the scattering channels are fully decoupled, and where the $\boldsymbol{\hat  l}^2$ operator is diagonal (with eigenvalues $\ell(\ell+1)$):
\begin{equation}
\label{eq:sfl}
{R}_{vj \ell,v'j' \ell'}^{\Gamma J} = \dfrac{\ell(\ell+1)}{2 m R^{2}} \delta_{vj\ell,v'j'\ell'}\,.
\end{equation}
Similarly to Eq.(\ref{eq:wftot}), the total wave function $\vert \Gamma ,J; E\rangle$ in the SF frame is expressed as
\begin{equation}
\label{eq:sfwftot}
\vert \Gamma ,J; E\rangle=\sum_{\alpha=1}^{N_{\textrm{tot}}} C_{\alpha}(E)\vert\Gamma,J,\alpha;E\rangle\,,
\end{equation}
where the channel wave functions $\vert \Gamma,J,\alpha; E\rangle$ are expanded on the basis set $\vert\Gamma,J,v,j,\ell\rangle$
\begin{equation}
\label{eq:sfwf}
\vert \Gamma,J,\alpha; E\rangle=\sum_{\kappa=1}^{N_{\textrm{tot}}} F_{\kappa,\alpha}(R;E) \vert\Gamma,J,v_{\kappa},j_{\kappa},\ell_{\kappa}\rangle\,.
\end{equation}
Like the channel index $a$ in the BF frame, the index $\alpha$ can be assigned at large distances to the asymptotic channel \textit{i.e.} $(\alpha \equiv v_{\alpha},j_{\alpha},\ell_{\alpha})$. The $F_{\kappa,\alpha}$ coefficients are solutions of Eq.~(\ref{eq:bfccset}) where the matrices are expressed in the SF basis set. The matrix elements of \textbf{V} in the SF frame are related to those in the BF frame according to
\begin{equation}
\label{eq:sfpot}
V_{vj\ell,v'j'\ell' }^{\Gamma J}(R)=\sum_{\Omega=0}^{J}\langle \ell \vert \Omega\rangle^{\Gamma J} V_{vj\Omega,v'j'\Omega }^{\Gamma J }(R)\langle\Omega \vert \ell'\rangle^{\Gamma J},
\end{equation}
where $\langle\ell \vert \Omega \rangle^{\Gamma J}$ is a matrix element of the transformation between BF and SF angular basis sets
\begin{equation}
\label{eq:ft}
\langle\ell \vert \Omega \rangle ^{\Gamma J}= (-1)^{J+\Omega}\sqrt{2\ell+1}\sqrt{2-\delta_{\Omega,0}}
\left(
\begin{array}{ccc}
 j&J&\ell \\
\Omega&-\Omega&0
\end{array}
\right)\,.
\end{equation}
The total parity $(-1)^J$ and $(-1)^{J+1}$ for $+$ (or \textit{e}) and $-$ (or \textit{f}) states imposes that the sum in Eq.(\ref{eq:sfwf}) is restricted to the $j$ and $\ell$ values satisfying $(-1)^{J}=(-1)^{j+\ell}$ and $(-1)^{J+1}=(-1)^{j+\ell}$, respectively.

At this step, the coefficients of channel superposition in Eq.~(\ref{eq:wftot}) are still unspecified. They  are determined by applying appropriate boundary conditions when the set of differential equations of Eq. (\ref{eq:bfccset}) is solved. Bound states are characterized by imposing the standard boundary conditions for each channel $\alpha$
\begin{equation}
\begin{array}{lll}
\label{eq:boundlim}
F_{\kappa,\alpha}(0;E_{t})&=0& \forall \kappa \in \left[1,N_{\textrm{tot}}\right] \\
F_{\kappa,\alpha}(R;E_{t})& \xrightarrow{R\rightarrow\infty}& 0~~~\forall \kappa \in \left[1,N_{\textrm{tot}}\right]\,. \\
\end{array}
\end{equation}
These quantization conditions are fulfilled only for discrete energies $E_{t}$, corresponding to a discrete vibrational level $v_t$ of H$_3^-$ characterizing the number of nodes (in the $R$ coordinate) of the radial wave function in the dominant channel $\alpha'_0$ for given $\Gamma$ and $J$. We used the renormalized Numerov method \cite{johnson78} to solve the coupled equations of Eq.(\ref{eq:bfccset}) written in the SF frame, so that the resulting superposition coefficients $C_{\alpha}^{v_t}$ ensure that the outward-propagated and inward-propagated wave functions are equal with identical derivatives at a well-chosen matching distance \cite{johnson78}. Equations (\ref{eq:sfwftot}) and (\ref{eq:sfwf}) are recast for the wave function of an H$_3^-$ bound level $\vert \Gamma,J ,v_t\rangle $ as
\begin{eqnarray}
\label{eq:boundwf}
\vert \Gamma,J ;\alpha'_0,v_{t}\rangle &=&\sum_{\kappa'=1}^{N_{\textrm{tot}}} \Phi_{\Gamma J \kappa'}^{\alpha'_0 v_{t}}(R) \vert\Gamma,J,v_{\kappa'},j_{\kappa'},\ell_{\kappa'}\rangle \,,\\
\Phi_{\Gamma J \kappa'}^{\alpha'_0 v_{t}}(R)&=&\sum_{\alpha'=1}^{N_{\textrm{tot}}} C_{\alpha'}^{\alpha'_0 v_{t}} F_{\kappa',\alpha'}(R;E_{t})\,.
\end{eqnarray}
As an illustration, the calculated energies of lowest H$_3^-$ bound levels for $J=0$ to 6 corresponding to the binding of para-H$_2(v=0,j=0)$ with H$^-$ are displayed in Table \ref{tab:para_bound_states}, labeled with $v_t$ and with the approximate quantum numbers $v,j,\ell$ of the corresponding dissociation threshold. This is justified by the weak interaction of the lowest channel ($v=0, j=0, \ell$) with the other ones, as already noticed in paper I. There is a single dominant component in the expansion of Eq.(\ref{eq:boundwf}). The present energies for $J=0$ are found in agreement within 0.5~cm$^{-1}$ with the ones obtained in paper I using a different integration method. Beyond $J=6$ no bound level can exist anymore below the dissociation threshold. Bound levels with energies below the ortho-H$_2(v=0,j=1)$+H$^-$ threshold are reported in Table \ref{tab:ortho_bound_states}. In the $C_{2v}$ framework above, these levels correspond to the binding of a para-(ortho)-H$_2$ with H$^-$, and all other bound levels, which are found at energies below thresholds with non-zero even (odd) values of $j$ correspond to predissociation resonances, which will decay into the continua of the H$_2(v=0,j')$+H$^-$ thresholds with $j'$even (odd) and $0 (1) \le j'<j$. All these levels are included in the RA cross section calculations below.

Note that such values could be used for the search for H$_3^-$ rotational transition lines in the ISM absorption spectra in the mm-wavelength range. As the component of the permanent electric dipole moment of H$_3^-$ along the $R$ axis exceeds by two orders of magnitude the transverse components (paper I), we can follow the notations of Ref.\cite{brown1975} for electric dipole transitions in a linear molecule. $Q$ lines ($\Delta J=0$) will occur between \textit{e} and \textit{f} states, and will not be observed between levels of Table \ref{tab:para_bound_states}. $P$ and $R$ lines ($\Delta J=\pm 1$) will connect levels with the same parity \textit{e} or \textit{f}.
\begin{table}[t]
\begin{tabular}{|p{1.5cm}|p{1.5cm}||p{1.5cm}|p{1.5cm}|}
\hline
 $J,v,j,\ell,v_t$&  Energy& $J,v,j,\ell,v_t$&  Energy\\
\hline
$ 0,0,0,0,0$  & -71.2 & $ 1,0,0,1,1$ &	-23.8	\\
$ 1,0,0,1,0$  &	-68.2 &	$ 2,0,0,2,1$ &	-20.2  \\
$ 2,0,0,2,0$  &	-62.4 &	$3,0,0,3,1$  &-15.0\\
$ 3,0,0,3,0$  & -53.8 & $4,0,0,4,1$  &  -8.6\\
$ 4,0,0,4,0$  & -42.8 & $5,0,0,5,1$  &  -1.7\\
$ 5,0,0,5,0$  & -29.7 & $0,0,0,0,2$  &	-5.4\\
$ 6,0,0,6,0$  & -15.3 & $1,0,0,1,2$  &	-4.5\\
$ 0,0,0,0,1$  &	-25.7 &	$2,0,0,2,2$  &	-2.7\\
$ ~~~~~    $  & ~~    &	$3,0,0,3,2$  &-0.4\\
\hline
\end{tabular}
\caption{Computed binding energies (in cm$^{-1}$) of the vibrational levels $v_t$ ordered as rotational progressions in $J$, of the para-H$_2$-H$^-$ complex with respect to the lowest dissociation limit H$_2(v=0,j= 0)+$H$^-$. Levels are labeled with the approximate quantum numbers $v=0,j=0$, and $\ell \equiv J$. In the super-dimer picture, they are all of \textit{e} parity.}
\label{tab:para_bound_states}
\end{table}
\begin{table}[t]
\begin{tabular}{|p{1.5cm}|p{1.5cm}||p{1.5cm}|p{1.5cm}|}
\hline
 $J,v,j,\ell,v_t$&  Energy& $J,v,j,\ell,v_t$&  Energy\\
\hline
$ 0,0,1,1,0$  & -152.7 & $ 0,0,1,1,2$  & -33.2 \\
$ 1,0,1,0,0$  &	-149.3 & $ 1,0,1,0,2$  & -31.6  \\
$ 2,0,1,1,0$  & -142.6 & $ 2,0,1,1,2$  &  -28.4 \\
$ 3,0,1,2,0$  & -132.6 & $ 3,0,1,2,2$  & -23.9\\
$ 4,0,1,3,0$  & -119.7 & $ 4,0,1,3,2$  &  -18.2\\
$ 5,0,1,4,0$  & -104.0 & $ 5,0,1,4,2$  &  -11.8 \\
$ 6,0,1,5,0$  &  -85.8 & $ 6,0,1,5,2$  &  -5.0  \\
$ 7,0,1,6,0$  &  -65.6 & $ 1,0,1,2,0$  & -13.3\\
$ 8,0,1,7,0$  &  -44.0 & $ 2,0,1,3,0$  &  -10.0 \\
$ 9,0,1,8,0$  &  -21.7 & $ 3,0,1,4,0$  &  -5.3 \\
$ 0,0,1,1,1$  & -77.8 &  $ 0,0,1,1,3$  & -10.7 \\
$ 1,0,1,0,1$  & -75.3 & $ 1,0,1,0,3$  & -9.8 \\
$ 2,0,1,1,1$  & -70.5 & $ 2,0,1,1,3$  &  -8.1 \\
$ 3,0,1,2,1$  & -63.3 & $ 3,0,1,2,3$  &  -5.6 \\
$ 4,0,1,3,1$  & -54.2 & $ 4,0,1,3,3$  &  -2.7 \\
$ 5,0,1,4,1$  & -43.2 & $ 0,0,1,1,4$  & -1.7 \\
$ 6,0,1,5,1$  & -31.0 & $ 1,0,1,0,4$  & -1.3 \\
$ 7,0,1,6,1$  & -17.9 & $ 2,0,1,1,4$  &  -0.7 \\
$ 8,0,1,7,1$  & -4.9  & $ 1,0,1,0,5$  & -0.07 \\
\hline
\hline
$ 1,0,1,1,0$  & -13.4 &$ 4,0,1,4,0$  & -0.5 \\
$ 2,0,1,2,0$  & -10.2 &$ 1,0,1,1,1$  & -0.1\\
$ 3,0,1,3,0$  & -5.8  &    ~&~\\
\hline
\end{tabular}
\caption{Computed binding energies (in cm$^{-1}$) of the vibrational levels $v_t$ of the ortho-H$_2$-H$^-$ complex ordered as rotational progressions in $J$, with respect to the lowest dissociation limit H$_2(v=0,j= 1)+$H$^-$). Levels are labeled with the approximate quantum numbers $v=0,j=1$, and $\ell$. In the super-dimer picture, levels in the upper part are of \textit{e} parity, and in the lower part of \textit{f} parity.}
\label{tab:ortho_bound_states}
\end{table}

In a scattering process with energy $E$, the number of energetically open channels $N_{o}$ gives the number of physical solutions of Eq.(\ref{eq:bfccset}). The remaining $N_{c}=N_{\textrm{tot}}-N_o$ closed channels should lead to non-physical long-range behavior since the related wave functions diverge asymptotically. Accordingly, the sum in Eq. (\ref{eq:sfwftot}) is restricted to the $N_o$ open channels
\begin{eqnarray}
\label{eq:scatwf}
\vert \Gamma,J,\alpha; E\rangle= \\ \sum_{\alpha=1}^{N_{o}}~C_{\alpha}(E) \sum_{\kappa=1}^{N_{\textrm{tot}}} F_{\kappa,\alpha}(R;E) \vert\Gamma,J,v_{\kappa},j_{\kappa},\ell_{\kappa}\rangle\,, \nonumber
\end{eqnarray}
and the usual boundary conditions are applied for each $\alpha$ to the set of coupled equations
\begin{equation}
\begin{array}{lll}
\label{eq:contlim}
F_{\kappa,\alpha}(0;E)&=0& \forall \alpha \in \left[1,N_{\textrm{tot}}\right] \\
F_{\kappa,\alpha}(R;E)& \xrightarrow{R\rightarrow\infty}& 0~~~\forall \alpha \in \left[N_o+1,N_{\textrm{tot}}\right] \\
F_{\kappa,\alpha}(R;E)& \xrightarrow{R\rightarrow\infty}& {j}(R,k_{\kappa}) \delta_{\kappa,\alpha}-{n}(R,k_{\kappa}) K_{\kappa,\alpha}(E)\\ &&~~~\forall \alpha \in \left[1,N_o \right]\,.
\end{array}
\end{equation}
The standard collision matrix \textbf{K}$(E)$ summarizes the channel interactions. The functions ${j}(R,k_{\kappa})$ and ${n}(R,k_{\kappa})$ are the  regular and irregular spherical Bessel functions with momentum $k_{\kappa}$ with respect to the H$_2$($v_{\kappa},j_{\kappa},\ell_{\kappa}$)+H$^-$ limit. To be fully determined, the total wave function $\vert \Gamma,J,E\rangle$ must also specify the initial state of the scattering process, \textit{i.e.} the entrance channel $\alpha_0$, through standard outgoing wave normalization
\begin{equation}
\label{eq:cscat}
C_{\alpha}^{\alpha_{0}}(E)= \left[\textbf{I}-\imath \textbf{K}(E)\right]^{-1}_{\alpha,\alpha_{0}}
\end{equation}
\section{Cross-section for dimer-atom radiative association}
\label{sec:RA-theory}

The RA of H$^-$ with H$_2$ in an initial rovibrational level $(v,j)$, creating H$_3^-$ in a vibrational level $v_t$, with respect to a dissociation limit H$^-$+H$_2$($v',j'$)
\begin{equation}
\label{eq:ra}
\mathrm{H}_2(v,j)+\mathrm{H}^- \to  \mathrm{H}_3^-(\Gamma',J',v_t (v',j')) +\hbar\omega_t\,,
\end{equation}
is schematically shown in Fig. \ref{fig:ad_curves} for $v=0,j=0$. A similar picture holds for H$^-$ and ortho-H$_2(v=0,j=1)$ association. We limit our study to these two states, which are the only ones populated in the cold and dense molecular ISM (see, for instance, Ref. \cite{rachford2009}). The two partners approach each other with a collision energy $E$ above the threshold. A photon with energy $\hbar\omega_{t}=E+E_{v_t}$ is emitted to stabilize H$_3^-$ in the vibrational level $v_t$ with binding energy $E_t$, with respect to the H$_2(v=0,j=0)+$H$^-$ or the H$_2(v=0,j=1)+$H$^-$ limits.

In order to obtain the RA cross-section and rate coefficient, we modify the theoretical approach developed by Herzberg \cite{herzberg50}, which was later used by several authors \cite{zydelman90,stancil93,gianturco97} for RA in diatomic molecules, and for photo-association of cold atoms \cite{cote96,pillet97}. This formalism is somewhat similar to the treatment one of us used previously for the photodissociation of van der Waals systems, which can be considered as the inverse process \cite{fourre1994}. The initial and final states are specified in the SF frame by $|\Gamma,J,E,\alpha_{0}\rangle$ and $|\Gamma,J,v_t,\alpha'_{0}\rangle$. The entrance channel index $\alpha_0$ is correlated to the asymptotic quantum numbers $v_0,j_0$ of the initial H$_2$ level, and $\ell_0$ specifying the initial collisional state. The final state of the created bound H$_3^-$ molecule has a multichannel nature, but it is labeled for convenience with the vibrational index $v_t$, and the index $\alpha'_{0}$ of the dominant channel in its expansion (Eq.(\ref{eq:boundwf})), also related to the asymptotic quantum numbers $\alpha'_0 \equiv v'_0,j'_0,\ell'_0$ (as in Tables \ref{tab:para_bound_states} and \ref{tab:ortho_bound_states}).

The Einstein coefficient $A^{\Gamma'J',\Gamma J}_{\alpha'_0;\alpha_{0}}(E,v_t)$ for the process of Eq. (\ref{eq:ra}) is written (in atomic units) \cite{gianturco97}
\begin{eqnarray}
\label{eq:adef}
A^{\Gamma'J';\Gamma J}_{\alpha'_0;\alpha_{0}}(E,v_t)=\frac{4\omega_{t}^3}{3c^3}
 \sum_{\sigma=0,\pm 1} \biggr| \langle \Gamma' J' v_t \alpha'_0\rvert \mu^{\sigma}\lvert \Gamma J E \alpha_0 \rangle \biggr| ^2\,.
\end{eqnarray}
%
%
where $\mu^{\sigma}=\mu^{0},\mu^{\pm 1}$ are the three components of the permanent dipole moment operator $\widehat{\boldsymbol{\mu}}$ in the SF frame.
The $\mu^{\sigma}$ components in the BF frame are related to the ones in the SF frame (determined in paper I) by the Wigner rotation functions $D$ depending on the Euler angles $\alpha_e$ and $\beta_e$
\begin{equation}
\mu^{\sigma}=\sum_{\lambda=0,\pm 1}\left[{D}_{\sigma,\lambda}^{1}(\alpha_e,\beta_e,0)\right]^*\mu^{\lambda}.
\end{equation}
We obtain
\begin{eqnarray}
\label{eq:adef1}
A^{\Gamma'J',\Gamma J}_{\alpha'_0;\alpha_{0}}(E,v_t)=\frac{4\omega_{t}^3}{3c^3} (2J+1)(2J'+1) \times \nonumber \\
\sum_{\sigma=0,\pm 1} \left(
\begin{array}{ccc}
 J& 1 & J' \\
M &\sigma  & -M-\sigma
\end{array}
\right)^2 \times \nonumber\\
\biggr|\sum_{\lambda=0,\pm 1}\langle \Gamma', J' ;v_t ,\alpha'_0\rvert \mu^{\lambda}\lvert \Gamma, J ;E ,\alpha_0 \rangle \biggr| ^2\,,
\end{eqnarray}
which becomes, after averaging over the $M$ values
\begin{eqnarray}
\label{eq:adef2}
A^{\Gamma'J',\Gamma J}_{\alpha'_0;\alpha_{0}}(E,v_t)=\frac{4\omega_{t}^3}{3c^3} (2J'+1) \times \nonumber \\
\biggr|\sum_{\lambda=0,\pm 1}\langle \Gamma' ,J'; v_t ,\alpha'_0\rvert \mu^{\lambda}\lvert \Gamma ,J; E ,\alpha_0 \rangle \biggr| ^2\,,
\end{eqnarray}
We now introduce the expressions for the total wave functions (Eqs. (\ref{eq:boundwf}, \ref{eq:scatwf}))
\begin{eqnarray}
\label{eq:adef2}
A^{\Gamma'J',\Gamma J}_{\alpha'_0;\alpha_{0}}(E,v_t)=\frac{4\omega_{t}^3}{3c^3} (2J'+1) \biggr|\sum_{\lambda=0,\pm 1} \nonumber \\
\sum_{\alpha=1}^{N_{o}}~C_{\alpha}^{\alpha_{0}}(E) d^{\Gamma' J',\Gamma J}_{\lambda \alpha}(E,v_t,\alpha'_0) \biggr| ^2\,,
\end{eqnarray}
with
\begin{eqnarray}
\label{eq:dipofr}
d^{\Gamma' J',\Gamma J}_{\lambda \alpha}(E,v_t,\alpha'_0)= \\
\sum_{\kappa'=1}^{N'_{\textrm{tot}}} \sum_{\kappa=1}^{N_{\textrm{tot}}} \langle \Phi_{\Gamma J \kappa'}^{\alpha'_0 v_{t}}(R) \rvert \mu_{\kappa',\kappa}^{\lambda}(R)\lvert F_{\kappa,\alpha}(R;E)\rangle_R. \nonumber
\end{eqnarray}
The subscript $R$ at the angled brackets denotes the integration over the $R$ coordinate \footnote{Note that in the numerical implementation, the initial (energy-normalized) and final (unity-normalized) wave functions are not directly computed with the renormalized Numerov method; the integral is progressively built during the integration process, as explained in the appendix of Ref. \cite{fourre1994}.}. The matrix elements of the $R$-dependent transition dipole moment in the SF basis are
\begin{eqnarray}
\label{eq:sfmu}
\mu_{\kappa',\kappa}^{\lambda}(R)=\\
\sum_{\Omega'=0}^{J'} \sum _{\Omega=0}^{J}\langle\ell'|\Omega' \rangle^{\Gamma' J'}\mu_{v'j'\Omega',vj\Omega}^{\lambda}(R)\langle\Omega|\ell \rangle^{\Gamma J}\,, \nonumber
\end{eqnarray}
related to the matrix elements in the BF basis
\begin{eqnarray}
\label{eq:bfmu}
\mu_{v'j'\Omega',vj\Omega}^{\lambda}(R)= (-1)^{\Omega'}
\left(
\begin{array}{ccc}
 J& 1 & J' \\
\Omega &\lambda  & -\Omega'
\end{array}
\right) \times \nonumber\\
 \langle \chi_{v'j'}(r) \Theta ^{\Omega'}_{j'} (\theta)| \mu^{\lambda}(R,r,\theta)|\chi_{vj}(r) \Theta ^{\Omega}_{j}(\theta)\rangle_{r,\theta}.
\end{eqnarray}
The subscripts ${r,\theta}$ at the angled brackets denote the integration over the $r$ and $\theta$ variables. Note that the initial and final wave functions are not necessarily described with the same number of channels, i.e. $N_{\textrm{tot}}$ and $N'_{\textrm{tot}}$ can be different because $J\ne J$ and $\Gamma\ne\Gamma'$. We use energy normalization for the initial continuum state.

\begin{figure}
 \includegraphics[width=7.cm]{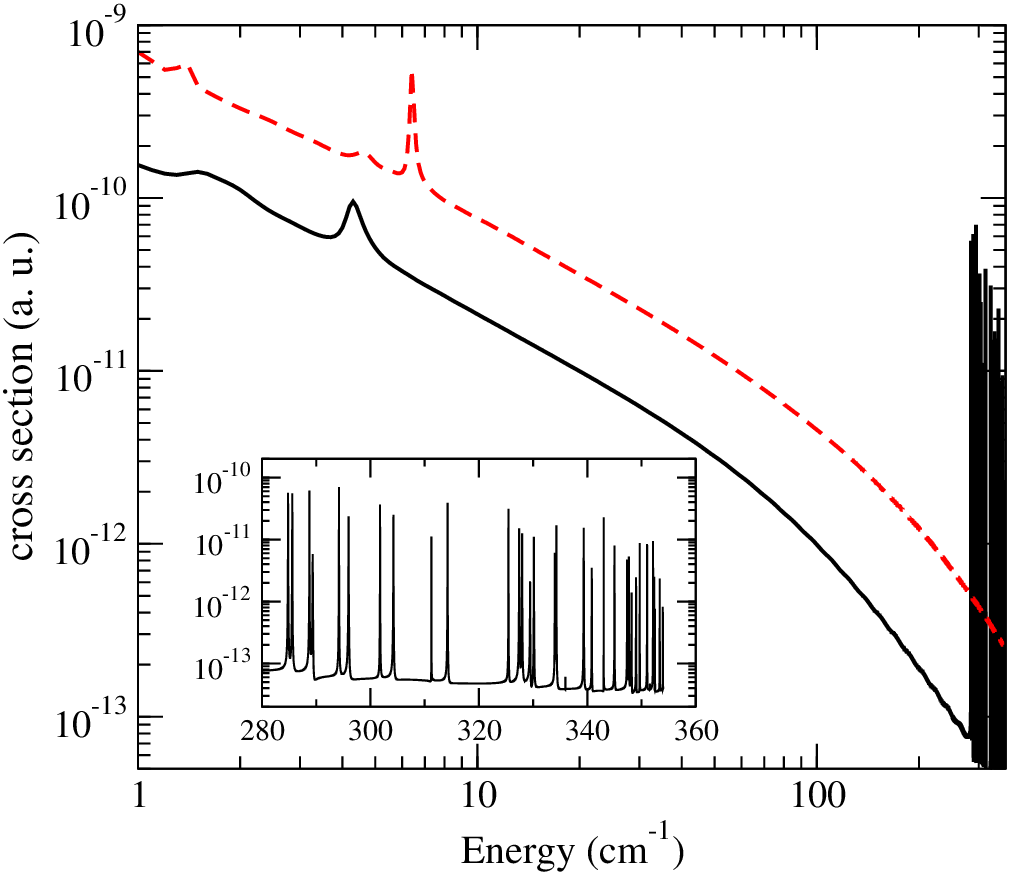}
 \caption{The RA cross-section (in atomic units) starting from  para-H$_2$ ($v_0=0, j_0=0$) (full line), and from  ortho-H$_2$ ($v_0=0, j_0=1$) (dashed line) as a function of collision energy $E$ above the corresponding thresholds. The inset enlarges the region of Feshbach resonances induced by the bound states of the closed channel para-H$_2$($v=0, j=2$)+H$^-$ (see Fig. \ref{fig:ad_curves}).}
 \label{fig:cs}
\end{figure}

The probability $P^{\Gamma'J',\Gamma J}_{\alpha'_0;\alpha_{0}}(E,v_t)$ of an RA event starting in the $|\Gamma,J;E,\alpha_{0}\rangle$ state with momentum $k=\sqrt{2mE}$ toward the final state $|\Gamma',J';v_t,\alpha'_0\rangle$ is given by the Einstein coefficient divided with the current density of incident particles, which is $1/(2\pi)$ for the energy normalized wave function. The corresponding RA cross-section is then expressed as
\begin{eqnarray}
\label{eq:cspartial0}
\sigma^{\Gamma'J',\Gamma J}_{\alpha'_0;\alpha_{0}}(E,v_t)=\pi P^{\Gamma'J',\Gamma J}_{\alpha'_0;\alpha_{0}}(E,v_t)/k^2\,.
\end{eqnarray}
In the following, the indexes $\alpha_0$ and $\alpha'_0$ are assigned to their asymptotic labeling $(v_0, j_0, \ell_0)$ and $(v'_0, j'_0, \ell'_0)$. Physically, the entrance channel is determined asymptotically by the initial rovibrational H$_2$ level $(v'_0, j'_0)$, so that a sum over $\ell_0$ must be performed. Since the nuclear spin is conserved during the RA process, there is no need to include the nuclear spin degeneracy factor. All possible final states should also be included for the computation of the total RA cross section. For each initial $J$ value, the contributions from each final value $J'=J,J\pm 1$ must be added together. At a given collision energy $E$, the summation over all possible values of $J$ must be performed as well. Finally, we are interested in the formation of H$_3^-$ in any of its stable bound levels ${v_t}$, so that the total cross section is obtained as follows
\begin{eqnarray}
\label{eq:cstot}
\sigma_{v_{0}j_0}(E)=\sum_{\Gamma'J'\Gamma,J v'_{t} \ell_0 v'_0, j'_0, \ell'_0} \sigma^{\Gamma'J',\Gamma J}_{v'_0, j'_0, \ell'_0;v_0, j_0, \ell_0}(E,v_t)\,.
\end{eqnarray}
As reported in paper I, the component $\mu^0(R,r,\theta)$ (along $Z$-axis) of the H$_3^-$ dipole moment in Eq.~(\ref{eq:bfmu}) is by two orders of magnitude larger than the two others components $\mu^{\pm 1}(R,r,\theta)$. Therefore, the contribution from the transverse components is neglected. The total cross sections $\sigma_{00}(E)$ and $\sigma_{01}(E)$ for the RA of H$^-$ with para-H$_2$ ($v=0, j=0$) and with ortho-H$_2$ ($v=0, j=1$) respectively, are shown in Fig. \ref{fig:cs}. We already mentioned that at low collision energies (below 400~cm$^{-1}$), a single component dominates the multichannel expansion of the initial and final wave functions. Therefore, the approximate selection rules $v_i \to v_f=v_i, j_i \to j_f=j_i$, and $J \to J'=J\pm 1$ hold for the RA process (with the appropriate selection rule for the parity). As expected from Eq.(\ref{eq:adef2}), the cross section for RA with ortho-H$_2$ is larger than the one with para-H$_2$, as the former species has a deeper potential well, and has more bound states than the latter. The bumps near 1.5~cm$^{-1}$ visible in both curves are due to the enhancement of the probability density at the top of the centrifugal barrier for low $J$. Moreover, one shape resonance is predicted near 4.3~cm$^{-1}$ in $\sigma_{00}(E)$ and near 6.5~cm$^{-1}$ in $\sigma_{01}(E)$ associated with non-zero $J$ values.  Once the collision energy reaches  285~cm$^{-1}$ at which the $j=2$ rotational state of para-H$_2$ can be populated, a series of Feshbach resonances induced by the bound states of the closed channel para-H$_2$($v=0, j=2$)+H$^-$ appears in the cross section.

\begin{figure}
 \includegraphics[width=7.cm]{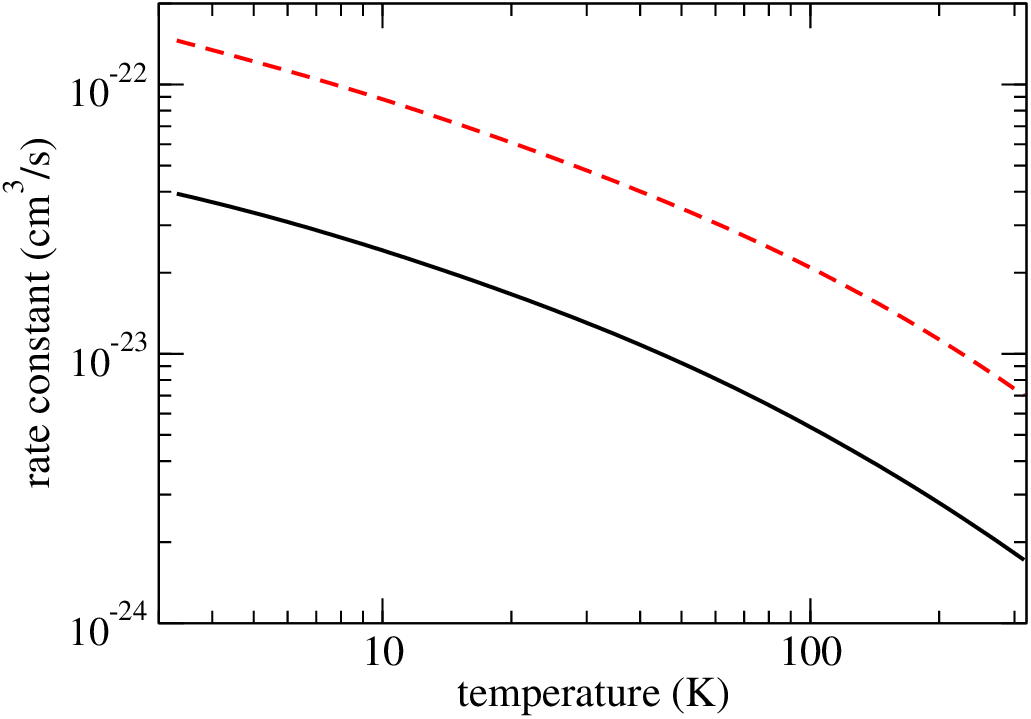}
 \caption{The rate coefficient $k_{\rm RA}$ as a function of temperature for RA of H$^-$ with para-H$_2$ ($v_0=0, j_0=0$) (full line) and with ortho-H$_2$ ($v_0=0, j_0=1$) (broken line).}
 \label{fig:rate}
\end{figure}

The RA rate coefficient $k_{\rm RA}(T)$ is obtained by a standard integration over the Maxwell-Boltzmann collision velocity distribution. Its variation with temperature is displayed in Fig. \ref{fig:rate} for the RA of H$^-$ with para-H$_2$ ($v_0=0, j_0=0$) and with ortho-H$_2$ ($v_0=0, j_0=1$). As a follow-up of Fig. \ref{fig:cs}, the rate is found about 4 times larger in the ortho-H$_2$ case than in the para-H$_2$ case. Note that the resonances in the latter case are too narrow to influence the rate above 200~K. The relative abundances of ortho-H$_2$ and para-H$_2$ could be out of thermal equilibrium in the interstellar clouds \cite{rachford2009}. The obtained RA rates suggest that the ratio of ortho-H$_2-$H$^-$ to para-H$_2-$H$^-$" is enhanced by a factor of 4 compared to the ratio of ortho-H$_2$ to para-H$_2$.

\section{Discussion: is the formation of H$_3^-$ in the ISM possible?}
\label{sec:ism}

The production rate of H$_3^-$ in the ISM is directly related to the presence of H$^-$. Here, we propose a rough estimation of the abundance of H$^-$ in the ISM assuming that it is formed only by dissociative attachment of an electron to H$_2$. Other possible mechanisms such as radiative attachment to H are neglected.
The chemistry of interstellar clouds is initiated by ionization of molecular hydrogen by cosmic rays with a typical rate constant  $\zeta\sim3\times  10^{-17}$s$^{-1}$ in diffuse interstellar clouds \cite{oka06b}.  (Cloud densities are $\sim 10^2$ cm$^{-3}$ in diffuse clouds and  $\sim 10^4$ cm$^{-3}$ in dense clouds.) The ionized molecular hydrogen H$_2^+$ quickly forms H$_3^+$ in collisions with H$_2$, with a rate constant  $\sim 2\times 10^{-9}$cm$^3$/s \cite{oka06b}. The electron escaped after ionization of H$_2$ has a large kinetic energy and undergoes many elastic collisions with environmental H$_2$ before it thermalizes. In each elastic collision with H$_2$, the electron looses a fraction (about $4m_e/m_{{\rm H}_2}$) of its incident energy. For example, the electron should experience about 6200 elastic collisions with H$_2$ before its energy is decreased from 10~eV to 10~meV ($T\sim$120~K).  Possible inelastic $e^-$+H$_2$ collisions will lead to vibrational excitation of H$_2$ and to dissociative attachment (DA), $e^-$+H$_2\to$ H+H$^-$. The DA reaction is allowed for collision energies above the threshold at 3.7 eV with the cross section  about $\sigma_{DA}\sim 3\times 10^{-21}$~cm$^2$ at 4 eV \cite{schulz65,horacek04}.

We can roughly estimate  the fraction  $f_{\rm DA}$ of escaped electrons that would form H$^-$ rather than undergo the thermalization process. This gives us an estimate about the production rate of H$^-$ in the ISM. The upper bound for the $e^-+$H$_2$ elastic cross-section $\sigma_{el}$ is about $1.7\times 10^{-15}$ cm$^2$ \cite{furst84}. If we take $N_{th}=1000$ elastic collisions corresponding to the thermalization down to the energy below which DA is impossible, we should compare $\sigma_{el}/N_{th}$ with $\sigma_{DA}$. It gives approximately the value $f_{\rm DA}=\sigma_{DA}N_{th}/\sigma_{el}\sim 0.0018$ for the fraction of escaped electrons that form H$^-$ by dissociative attachment to H$_2$.  Therefore, the rate of H$^-$ production in cm$^3$ in the ISM can be estimated as $\zeta n(\mathrm{H}_2) f_{\rm DA}=\zeta n(\mathrm{H}_2)\sigma_{DA}N_{th}/\sigma_{el}$. The H$^-$ ion has only one bound electronic state and cannot be detected directly. However, if it forms a molecular ion AH$^-$ by an association with an atom or molecule A, its presence in the ISM can be proven indirectly by rovibrational absorption  spectroscopy of AH$^-$.

One of the motivation for this study was the investigation whether H$_3^-$ can be formed in the ISM and be detected by photoabsorption spectroscopy. The stability of H$_3^-$ was confirmed in this study. Therefore, H$_3^-$ can exist in the ISM in cold clouds ($T<100$~K). However, to be observed by far-infrared absorption spectroscopy, H$_3^-$ should be present in the ISM in relatively large amounts.  The H$_3^-$ molecular ion and the H$^-$ ion can be destroyed in the ISM by cosmic rays and by mutual neutralization with positive ions. These processes limit the absolute abundance of H$_3^-$. In order to determine the H$_3^-$ abundance, one has to consider rate equations for all reactions involving formation and removal of H$^-$ and H$_3^-$ in the ISM. However, a rough estimation about H$^-$ and H$_3^-$ abundance in the ISM can be made using the available data.

First, we estimate the H$^-$ abundance $n({\rm H^-})$. According to our model the rate of H$^-$ production in 1~cm$^3$ is given by $\zeta n({\rm H}_2)f_{\rm DA}$. The principal channel of destruction is probably due to the mutual neutralization with positive ions. The number density $n_{+}$ of positive ions can be taken to be equal to the number density of electrons in the ISM. In diffuse clouds, the electron number density is about 0.1  \% of $n({\rm H}_2)$, i.e. $n_{+}\sim0.01$ cm$^{-3}$. Therefore, the rate of removal of H$^-$ from the ISM is $k_\pm n({\rm H^-}) n_{+}$ (in 1 cm$^3$), where $k_\pm $ is the rate coefficient for the mutual neutralization. We take the value of $k_\pm \sim 10^{-7}$cm$^3/$s for the H$^-$+H$^+\to$H+H reaction \cite{bates55,peart92,janev06} at 10 meV. Therefore, we derive the equilibrium abundance of H$^-$: $n({\rm H^-})=\zeta n({\rm H}_2)f_{\rm DA}/(k_\pm   n_{+})\sim 5\times10^{-7}$cm$^{-7}$. If we take the size of a cloud to be 10 pc, the resulting column density is $\sim 10^{13}$ cm$^{-2}$, which is a reasonable number for observation.

The estimation of the H$_3^-$ abundance $n({\rm H_3^-})$ can be made in a similar way. The rate of H$_3^-$ removal is determined by the similar formula: $k_\pm n({\rm H_3^-}) n_{+}$. The rate of H$_3^-$ formation is $k_{\rm RA}n({\rm H^-})n({\rm H_2})$. The equilibrium H$_3^-$ abundance is then given by  $n({\rm H_3^-})=k_{\rm RA}n({\rm H^-})n({\rm H_2})/(k_\pm n_{+})\sim 10^{-17}$cm$^{-3}$. Such an abundance would produce a column density, which is  small even for a large interstellar cloud.

If we combine the formulas for H$^-$ and H$_3^-$ abundances we obtain
\begin{equation}
 n({\rm H_3^-})=\frac{k_{\rm RA}\zeta f_{\rm RA}}{(k_\pm)^2 }\left(\frac{n({\rm H_2})}{n_{+}}\right)^2\,.
\end{equation}
The formula suggests that the H$_3^-$ abundance should be larger in an environment with smaller degree of ionization (i.e. with a smaller number density of positive ions). The environment should be cold enough ($<$100~K) for the H$_3^-$ ion to be stable with respect to collisions with other species. Therefore, if H$_3^-$ can be detected in the ISM, one has to search for it in cold dense interstellar clouds. If H$_3^-$ is detected it would be a proof that H$^-$ is also present in the ISM.

\section*{Acknowledgments} We thank Roland Wester for motivating us to study H$_3^-$ structure and dynamics. Stimulating discussions with Prof. Jacques Robert are gratefully acknowledged. The study was supported by Triangle de la physique as part of the project "Quantum Control of Cold Molecules" (contract QCCM-2008-007T), and by the National Science Foundation under grant PHY-0855622.

%

\end{document}